\documentclass[journal]{IEEEtran}
\usepackage{amsmath,amsthm,amsfonts,amssymb}
\usepackage{tikz}
\usepackage{url}
\usepackage[nolist]{acronym}
\usepackage[colorlinks]{hyperref}
\hypersetup{
colorlinks,
    linkcolor={red!40!black},
    citecolor={blue!40!black},
    urlcolor={blue!20!black}
}

\graphicspath{{./Figures/}}
\usepackage{commands}

\begin{document}

\title{Capacity Bounds and Degrees of Freedom for MIMO Antennas Constrained by Q-Factor}

\author{Casimir~Ehrenborg,~\IEEEmembership{Member,~IEEE}, Mats~Gustafsson,~\IEEEmembership{Senior Member,~IEEE}, \\and Miloslav~Capek,~\IEEEmembership{Senior Member,~IEEE}
\thanks{Manuscript received \today. This work was supported by the Swedish foundation for strategic research (SSF) under the program applied mathematics and the project Complex analysis and convex optimization for electromagnetic design and by the Czech Science Foundation under Project 19-06049S.}%
\thanks{C. Ehrenborg is with the Department of Electrical Engineering, KU Leuven, Kasteelpark Arenberg 10 postbus 2440, 3001 Leuven, Belgium. (E-mail: casimir.ehrenborg@kuleuven.be).}
\thanks{M. Gustafsson is with the Department of Electrical and Information Technology, Lund University, Box 118, SE-221 00 Lund, Sweden. (Email: mats.gustafsson@eit.lth.se).}
\thanks{M.~Capek is with the Department of Electromagnetic Field, Faculty of Electrical Engineering, Czech Technical University in Prague, 166~27 Prague, Czech Republic (E-mail: \mbox{miloslav.capek@fel.cvut.cz}).}
}

\maketitle

\begin{abstract}
The optimal spectral efficiency and number of independent channels for MIMO antennas in isotropic multipath channels are investigated when bandwidth requirements are placed on the antenna. By posing the problem as a convex optimization problem restricted by the port Q-factor a semi-analytical expression is formed for its solution. The antennas are simulated by method of moments and the solution is formulated both for structures fed by discrete ports, as well as for design regions characterized by an equivalent current. It is shown that the solution is solely dependent on the eigenvalues of the so-called energy modes of the antenna. The magnitude of these eigenvalues is analyzed for a linear dipole array as well as a plate with embedded antenna regions. The energy modes are also compared to the characteristic modes to validate characteristic modes as a design strategy for MIMO antennas. The antenna performance is illustrated through spectral efficiency over the Q-factor, a quantity that is connected to the capacity. It is proposed that the number of energy modes below a given Q-factor can be used to estimate the degrees of freedom for that Q-factor.
\end{abstract}


\section{Introduction}
\label{sec:intro}

\IEEEPARstart{M}{odern} communication systems often make use of \ac{MIMO} antennas. They utilize spatial multiplexing to dramatically increase the transmitted bit-rate, or capacity, in comparison to single antenna systems~\cite{Paulraj+etal2003,Molisch2011}. The optimal performance of \ac{MIMO} systems is therefore a topic of great interest. Many different methods have been developed to evaluate maximum capacity under different conditions and assumptions. For example by considering spherical regions~\cite{AlayonGlazunov+etal2011,Gustafsson+Nordebo2007a}, or treating the problem through an information theoretical approach~\cite{Migliore2019,Migliore2008,Franceschetti+etal2009,Taluja+Hughes2012,Kundu2016}. In~\cite{Ehrenborg+Gustafsson2018,Ehrenborg+Gustafsson2020} the authors presented a novel method for calculating optimal spectral efficiency of arbitrary antenna designs using convex current optimization.

Design of \ac{MIMO} antennas is based on effectively exciting discrete communication channels with low correlation~\cite{Paulraj+etal2003}. A proposed strategy for accomplishing this is to design the antennas such that they effectively excite modes with orthogonal radiation patterns, such as characteristic modes~\cite{Manteuffel+Martens2011,Miers+etal2013,Li+etal2014,Alroughani+etal2016}. A modal analysis method for analyzing \ac{MIMO} antennas design, constrained by radiation efficiency, was presented in~\cite{Ehrenborg+Gustafsson2020}. This method is based on analyzing the eigenvalues of a set of modes known as radiation modes to predict the performance of a \ac{MIMO} antenna. However, one of the open questions of this technique is if it can be generalized to other constraining parameters, such as bandwidth.

Electrically small antennas suffer from a degradation in possible performance as their size is reduced compared to the wavelength. Some of the parameters where this is most evident are radiation efficiency, directivity, and bandwidth~\cite{Hansen2006,Volakis+etal2010,Wheeler1975,Gustafsson+etal2015b}. Bandwidth can be estimated, for electrically small systems, through the Q-factor defined as the quotient of energy stored in a system over power dissipated by it~\cite{Schab+etal2018}. This relation accurately estimates the bandwidth, for single feed, single resonance systems~\cite{Yaghjian+Best2005}, however, it does not hold for multi-port systems. In this paper, this is addressed by considering a multi-port system as a superposition of single port systems. Each port has a well defined Q-factor calculated from the stored energy of the current induced by that port~\cite{Vandenbosch2010}. These port currents make up the total current distribution on the antenna when added together. The Q-factor of the total current is therefore an implicit indication of the Q-factor in each port. In this way, restricting the Q-factor of the total current density implicitly imposes a bandwidth requirement on the system.

In this paper optimal spectral efficiency is calculated using the method from~\cite{Ehrenborg+Gustafsson2020} for \ac{MIMO} antennas transmitting in a uniform multipath environment restricted by a required Q-factor. The optimization problem is formulated in the sources of the antenna design problem, either the feeding voltage of a fixed antenna array, or the current distribution in a design region. These two cases are studied to gain different insights into how to maximize spectral efficiency.

The antenna array is optimized by recasting the optimization problem in the port voltages. This is used to study the placement of canonical dipole antennas to gain an intuition for maximizing spectral efficiency. An advantage of this approach is the feasibility of the bound via direct realization of the port voltage excitation proposed by the optimization routine. The current optimization problem is solved for a few sub-regions embedded in a plate. The performance and placement of these sub-regions are compared to optimizing the current over the entire plate.

The performance is measured in a new quantity, spectral efficiency over Q-factor, that is related to the capacity of the \ac{MIMO} system by multiplication with the center frequency. This quantity illustrates that the Pareto type bound between the spectral efficiency and Q-factor has an optimal point.

The modal analysis method in~\cite{Ehrenborg+Gustafsson2020} is generalized to \ac{MIMO} antennas restricted by the Q-factor by considering a different set of modes called energy modes~\cite{Gustafsson+etal2016a}.  It is shown that the energy modes can predict the onset of the solution, as well as the number of degrees of freedom utilized in the system. The energy modes are compared to the characteristic modes and it is shown that they have similar performance, validating previous design techniques. It is also demonstrated, through simulation, that a uniform Rayleigh channel does not significantly alter the behavior of the optimal spectral efficiency problem.

The paper is organized in the following way.
In Section~\ref{sec:theory} the full-wave \ac{MIMO} system is introduced and the effect of the channel on systems with few ports is discussed. The problem is stated and solved for port quantities in Section~\ref{sec:ports} and for current density in Section~\ref{sec:currents}. The paper is concluded in Section~\ref{sec:conclusions}.

\section{Full-wave Model of MIMO Systems}
\label{sec:mimo}

\begin{figure}
    \centering
    \includegraphics[width=0.8\linewidth]{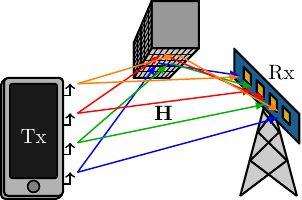}
    \caption{Schematic illustration of a \ac{MIMO} system, with a channel matrix~$\Hm$ modeling wave propagation from transmitter~Tx to receiver~Rx.}
    \label{fig:MIMO}
\end{figure}
\label{sec:theory}
A \ac{MIMO} system, like the one shown in Fig.~\ref{fig:MIMO}, is described by
\begin{equation}
\ym = \Hm\xm + \nm ,
\label{eq:MIMO}
\end{equation}
where $\ym$~is an $M\times 1$ vector containing the received signals, $\Hm$~is an $M\times N$ matrix known as the channel matrix, $\xm$~is an $N\times 1$ vector containing the input signals, and $\nm$~is an $M\times 1$ vector containing the noise perturbing each of the receivers~\cite{Paulraj+etal2003}. The channel matrix describes the structure of the antenna and wave propagation between the ports of the transmitter and the ports of the receiver. There are several ways to model a \ac{MIMO} system depending on how we wish to describe the input and output signals. For example, we may consider the input signals to be the voltage sources in the antenna feeds or the wave forms sent through the transmission lines connected to the antenna ports. In this section of the paper we model the \ac{MIMO} system using a scattering matrix, and consider the input and output signals in terms of spherical mode coefficients~\cite{Hansen1988,AlayonGlazunov+etal2011,Gustafsson+Nordebo2007a}. 

An antenna system's radiating and receiving properties can be described by the scattering matrix~\cite{Hansen1988}, 
\begin{equation}
    \begin{pmatrix}
    \Sm_{\mrm{1,1}} & \Sm_{\mrm{1,2}} \\
    \Sm_{\mrm{2,1}} & \Sm_{\mrm{2,2}}
    \end{pmatrix}
    \begin{pmatrix}
    \vv \\
    \boldsymbol{\alpha}
    \end{pmatrix}
    =
    \begin{pmatrix}
    \wv \\
    \boldsymbol{\beta}
    \end{pmatrix} ,
    \label{eq:scatmat}
\end{equation}
where $\vv$ are the signals incoming to the ports of the antenna, $\wv$ are the outgoing signals from the antenna, $\boldsymbol{\alpha}$ are the incoming spherical waves, $\boldsymbol{\beta}$ are the outgoing spherical waves,  $\Sm_{\mrm{1,1}}$ is the reflection matrix with dimension $N\times N$, $\Sm_{\mrm{2,2}}$ is the matrix representing how the antenna couples incoming and outgoing spherical waves with dimension $\infty\times\infty$, $\Sm_{\mrm{2,1}}$ is the $\infty\times N$ matrix coupling the input signals to the antennas to the radiated spherical waves, and $\Sm_{\mrm{1,2}}$ is the $N\times\infty$ matrix coupling the incoming waves to the signals in the ports. The subscripts are used to denote that these matrices are parts of the scattering matrix. Here, we assume that our antenna is made of linear, passive, reciprocal materials, such as metals and dielectrics, making the antenna a reciprocal system. This means that the antenna behaves equivalently when receiving or transmitting~\cite{Hansen1988}.

\begin{figure}
    \centering
    \includegraphics[width=0.5\linewidth]{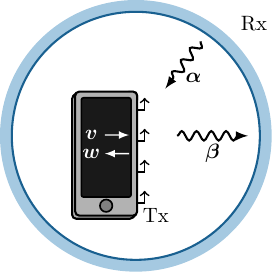}
    \caption{Schematic illustration of a \ac{MIMO} antenna modeled using a scattering matrix~\eqref{eq:scatmat}. The incoming and outgoing signals are depicted for the transceiving antenna as~$\vv$ and $\wv$, and for the spherical channel as $\boldsymbol{\alpha}$ and~$\boldsymbol{\beta}$, respectively.}
    \label{fig:sphMIMO}
\end{figure}

There exists many different channel models to describe specific propagation environments~\cite{Paulraj+etal2003,Molisch2011,Jakes+Cox1994}. However, in this paper we aim to consider the most general scenario to provide a basis from which to construct general performance bounds for \ac{MIMO} antennas. Many \ac{MIMO} devices, \eg handsets or small electronics, can be considered to operate in uniform multipath environments. These environments are characterized by a large number of independent waves impinging on the antenna from all directions, see Fig.~\ref{fig:sphMIMO}. These waves impinging on the sphere surrounding the antenna system are modeled by a Rayleigh channel in the spherical modes~\cite{AlayonGlazunov+etal2009,Gustafsson+Nordebo2007a}. The Rayleigh channel~$\Hmw$ is modeled by a random matrix with a complex Gaussian distribution that is uncorrelated and has zero mean~\cite{Paulraj+etal2003,Vaughan+Bach-Andersen2003,Miller1974}. Because the considered antenna is a reciprocal system it behaves equivalently when receiving or transceiving. Consequently, we model the transceiving channel as a Rayleigh channel as well.

The \ac{MIMO} system in Fig.~\ref{fig:MIMO} is therefore simplified to a \ac{MIMO} antenna in a spherical Rayleigh channel as in Fig.~\ref{fig:sphMIMO}. 
This system is governed by the equation
\begin{equation}
    \boldsymbol{\beta} = \Hmw\Sm_{2,1}\vv ,
\end{equation}
when operating in transceiving mode. The two parts of the channel model are here referred to as: the uniform Rayleigh fading channel, $\Hmw$, and the deterministic isotropic channel~$\Sm_{2,1}$. The instantaneous spectral efficiency of this system can be calculated by optimizing the time average over the modulation in the input signals, \ie $\Am= \medel{\vv\vv^{\herm}}/2$, where~$\medel{\cdot}$ denotes the temporal average. The matrix $\Am$ is called the covariance matrix of the input signals. The optimization problem for instantaneous spectral efficiency is~\cite{Paulraj+etal2003}
\begin{equation}
\begin{aligned}
& \underset{\Am\succeq\Om}{\maximize} && C =  \log_2\det(\Id+\gamma\Hmw\Sm_{2,1}\Am\Sm_{2,1}^{\herm}\Hmw^{\herm}) \\
& \subto && \Tr(\Am) = 1, \\
    \end{aligned}
    \label{eq:capacity}
\end{equation}
where $\Id$ is the identity matrix, input power is normalized to~$1$, and $\gamma$ is the \ac{SNR}. In many applications the channel $\Hmw$ changes over time. It is then necessary to consider the ergodic spectral efficiency, \ie the spectral efficiency averaged over channel realizations, $\bar{C} = \medel{C}$.

It was shown in~\cite{Gustafsson+Nordebo2007a} that the spectral efficiency for systems with high SNR can be decomposed into a deterministic part describing the antenna and a random part independent of the antenna, see Appendix~\ref{app:ray}. The physics of maximizing spectral efficiency are therefore independent of perturbing uniform channels, such as the Rayleigh channel. This is verified numerically in Section~\ref{sec:2dip_vardist}. We will therefore use the matrix~$\Sm_{2,1}$ as the channel matrix for our calculations ($\Hmw = \mat{1}$).

\begin{figure}
\centering
\includegraphics[width=7cm]{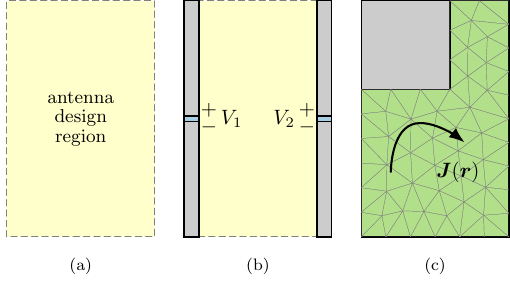}
\caption{Possible full-wave models of a transmitter Tx from Fig.~\ref{fig:sphMIMO}. (a) An antenna design region to be occupied by a MIMO system. (b) Two dipoles of fixed geometry are considered within the design region. The feeding voltages, confined to the blue regions, are controllable variables to be optimized for maximum capacity. (c) The current density, confined to the green region, is the controllable quantity. An optimized current is considered to be impressed in vacuum, the uncontrollable region is made from a given, yet arbitrary, material.}
\label{fig:portMIMO}
\end{figure}

Modelling a \ac{MIMO} system using a scattering matrix is useful when studying channel phenomena. However, in this paper, we focus on how the design of a \ac{MIMO} antenna impacts optimal spectral efficiency. Consider the region in Fig.~\ref{fig:portMIMO}a, we want to analyze how to construct the most effective \ac{MIMO} antenna within this region. We must therefore switch our modelling from scattering parameters to quantities that let us incorporate antenna design as a variable. 

The $\Sm_{1,1}$ part of the scattering matrix relation~\eqref{eq:scatmat} can be rewritten in terms of port quantities
\begin{equation}
    \zm\im = \vm,
    \label{eq:portImp}
\end{equation}
where $\vm$ are the port voltages, $\im$ are the port currents, and $\zm$ is the network impedance matrix~\cite{Pozar2005,Harrington1968}. Here, we have moved our perspective to Fig.~\ref{fig:portMIMO}b. In this case, the channel matrix~$\Sm_{2,1}$ is written as the connection between the port voltages and the outgoing spherical waves. We will denote this matrix~$\mat{s}$ for simplicity, defined in Section~\ref{sec:ports}.

The network impedance matrix can either be calculated from the (measured) scattering matrix or from a full-wave EM model. This is useful when studying a fixed set of antennas, as is done in Section~\ref{sec:ports}. We can further rewrite~\eqref{eq:portImp} in terms of meshed structures in order to use full-wave electromagnetic simulation techniques. In this paper we use \ac{MoM} to simulate our systems, rewriting~\eqref{eq:portImp} as,
\begin{equation}
    \Zm\Jm = \Vm ,
    \label{eq:MoM}
\end{equation}
where $\Jm$ is the current column matrix, $\Vm$ is the excitation column matrix, and $\Zm$ is the impedance matrix~\cite{Chew+etal2008}, see Appendix~\ref{app1}. This equation models the case in Fig.~\ref{fig:portMIMO}c, where the current can exist everywhere inside the design region. The current can be split into regions that are controlled or passively excited by the controlled region~\cite{Gustafsson+Nordebo2013}. Similarly to the port case, the channel matrix $\Sm_{2,1}$ is in this case written as the connection between the currents on the antenna to the modes in the far field and will be called matrix~$\Sm$ throughout the paper. This formulation is very useful when optimizing the current to study all possible antenna structures in a region, which is done in Section~\ref{sec:currents}.

\section{Array Antenna Excitation}
\label{sec:ports}

In this section we optimize the excitation of a fixed antenna geometry, see Fig.~\ref{fig:portMIMO}b, when the antenna is placed in a deterministic isotropic channel. The maximum capacity given by a set of feeding ports is studied when the antenna's bandwidth for each port is prescribed and represented by the Q-factor. These fixed geometries are used to convey an intuition of how to induce optimal spectral efficiency.

The current density on an antenna structure is modeled as a linear combination of currents excited by localized feeders, so-called port modes~\cite{Harrington+Mautz1972}, see Fig.~\ref{fig:portMIMO}b,
\begin{equation}
    \Jm = \Zm^{-1} \mat{C} \portv = \sum_p v_p \Jm_p,
    \label{eq:portMode1}
\end{equation}
where each port mode~$\Jm_p$ corresponds to a unitary excitation of the $p$-th port, $\Zm$ is the impedance matrix introduced in~\eqref{eq:MoM}, and~$\mat{C} \in \left\{0,1\right\}^{M \times P}$ is an indexation matrix defined element-wise as
\begin{equation}
C_{mp} = \left\{ {\begin{array}{*{20}{cl}}
1 & \textrm{$p$-th port is placed at $m$-th position,} \\
0 & \textrm{otherwise}.
\end{array}} \right.
\end{equation}
A remarkable advantage of the port-mode representation is the reduction of the original $M \times M$ matrices, \ie of number of degrees of freedom, to $P \times P$, the number of ports. The construction of the impedance matrix~\eqref{eq:MoM} prior the reduction into port modes~\eqref{eq:portMode1} is needed since it represents the full-wave electromagnetic behavior of the systems.  The values of physical quantities defined in \ac{MoM}-like~\eqref{eq:MoM} and port-like~\eqref{eq:portImp} representations are equivalent. To illustrate the transformation let $\mat{M} = \mat{M}^\herm \in \mathbb{C}^{M\times M}$ represent a generic \ac{MoM} operator, such as the reactance matrix~$\mat{X}$. The following quadratic forms map this operator to its port equivalent, $\mat{m}\in \mathbb{C}^{P\times P}$,
\begin{equation}
m = \dfrac{1}{2} \Jm^\herm \mat{M} \Jm = \dfrac{1}{2} \portv^\herm \mat{m} \portv,
\label{eq:portMode4}
\end{equation}
where
\begin{equation}
\mat{m} = \mat{C}^\herm \Zm^{-\mathrm{H}} \mat{M} \Zm^{-1} \mat{C}.
\label{eq:portMode3}
\end{equation}
Notice that the port-mode representation~\eqref{eq:portMode3} changes the physical units from Ohms for operator~$\mat{M}$ to Siemens for port operator~$\mat{m}$.

Spectral efficiency is calculated through the covariance of the input signals. The quadratic forms that typically calculate antenna quantities can be rewritten in terms of the covariance of the port voltages. Conveniently, the port voltages is the only part of the antenna expressions that vary in time, therefore the temporal average can be confined to this variable. For example the input power can be written as,
\begin{equation}
\begin{aligned}
    \frac{1}{2}\medel{\portv^\herm\mat{r}\portv} &= \frac{1}{2}\Tr\left(\medel{\portv^\herm\mat{r}\portv}\right) \\
    &= \frac{1}{2}\Tr\left(\medel{\mat{r}\portv\portv^\herm}\right) = \Tr(\mat{r}\mat{a}) = P_\mathrm{r},
\end{aligned}
\end{equation}
where $\mat{a}=\medel{\portv\portv^\herm}/2$ is the covariance of the port voltages, $P_\mathrm{r}$ is the radiated power, and the cyclic properties of the trace have been utilized. We assume the materials used in the antenna are lossless, therefore the radiated and input power is the same. This enables the real part of the impedance matrix, $\zm=\mat{r}+\ju\xm$, to be expanded in the spherical modes as $\mat{r}=\sm^\herm \sm$. The matrix~$\sm$ is the same matrix as the deterministic isotropic channel matrix, which can be calculated through~\eqref{eq:portMode3}, $\mat{s} = \Sm \Zm^{-1}\mat{C}$, see Appendix~\ref{app1} for the definition of matrix~$\Sm$~\cite{Tayli+etal2018,AlayonGlazunov+etal2011}.

Applying the change of basis~\eqref{eq:portMode4} and writing the problem in trace formulation, the spectral efficiency calculation~\eqref{eq:capacity} can be written in the port-mode basis as
\begin{equation}
\begin{aligned}
& \underset{\mat{a}}{\maximize} &&  \log_2\det\left(\Id+\gamma\mat{s}\mat{a}\mat{s}^{\herm}\right) \\
& \subto && \Tr(\mat{r}\mat{a})=1. 
\end{aligned}
\label{eq:portCapOpt}
\end{equation}
Here, the total radiated power is normalized to unity, $P_\mathrm{r}=1\unit{W}$. This problem is only restricted by the input power to each port.

Antennas are most often optimized given certain restrictions to their performance quantities, the formulation of~\eqref{eq:portCapOpt} enables us to add such restrictions as additional constraints. In this paper we investigate how bandwidth restrictions affect spectral efficiency optimization.

The Q-factor is an antenna quantity that estimates the fractional bandwidth~\cite{Yaghjian+Best2005,Schab+etal2018}, see its definition in Appendix~\ref{app:Q}. The connection between the Q-factor and the bandwidth is explicit for single input systems in free space but there exists no explicit quantity or expression for calculating the bandwidth for \ac{MIMO} systems. However, each of the ports of a \ac{MIMO} system is a single input system, if taken in isolation. Therefore, each of these ports have a well defined Q-factor based on their inputs. We can require a certain Q-factor of each of the ports in the system as a way of implicitly placing a requirement on the total systems bandwidth. However, the Q-factor is usually defined in relation to the system at resonance. For a multi-port system a single resonance of the total current is not always desired, thus the Q-factor of the system must be estimated in a different way. Here, we use the average of the magnetic and electric stored energies over the radiated power to estimate the bandwidth. This corresponds to adding the condition,
\begin{equation}
    \dfrac{\medel{\portv^{\herm}\mat{w}_\mathrm{x}\portv}}{2P_\mathrm{r}} = \frac{\Tr(\mat{w}_\mathrm{x}\mat{a})}{P_\mathrm{r}} \leq  Q,
    \label{eq:Qcond}
\end{equation}
where~$\mat{w}_\mathrm{x}$ is the matrix giving the reactive power of the averaged stored energy, see Appendix~\ref{app:Q}, and $Q$ is the required Q-factor.

The Q-factor constraint~\eqref{eq:Qcond} is added to~\eqref{eq:portCapOpt} to create the optimization problem that is investigated in this paper,
\begin{equation}
\begin{aligned}
& \underset{\mat{a}}{\maximize} &&  \log_2\det\left(\Id+\gamma\mat{s}\mat{a}\mat{s}^{\herm}\right) \\
& \subto && \Tr(\mat{r}\mat{a}) = 1, \\ 
& && \Tr(\mat{w}_\mathrm{x}\mat{a}) \leq  Q, \\
& && \mat{a} \succeq 0. \\
\end{aligned}
\label{eq:MIMOoptQport}
\end{equation}
 A problem of this form can be solved using commercially available software, such as \ac{CVX}~\cite{Grant+Boyd2011}, see~\cite{Ehrenborg+Gustafsson2018}. However, it can be solved much more efficiently, and in such a way as to provide greater physical insight, using the method presented in~\cite{Ehrenborg+Gustafsson2020}. The method is based on constructing a convex dual problem that can be solved semi-analytically. A brief outline of that procedure is covered here. 
 
Dual problems are constructed from an original optimization problem by forming linear combinations between the constraints of the original problem. The solution to the dual problem will therefore always have a value that bounds the solution of the original problem~\eqref{eq:MIMOoptQport}. However, there may exist a duality gap between the two solutions leading to a ``loose" bound on the optimization problem~\cite{Boyd+Vandenberghe2004}. The results presented in this paper have been verified numerically in CVX and no duality gap was observed.

To formulate the dual of the optimization problem~\eqref{eq:MIMOoptQport} we take an affine combination of the conditions restricting the original problem,
\begin{equation}
\begin{aligned}
& \min\limits_\nu\,\max\limits_{\mat{a}} &&  \log_2\det\left(\Id+\gamma\mat{s}\mat{a}\mat{s}^{\herm}\right) \\
& \subto && \Tr\left[\left(\nu\mat{r} +Q^{-1}\mat{w}_\mathrm{x}\right)\mat{a}\right]=(\nu+ 1)  , \\ 
& && \mat{a} \succeq 0. \\
\end{aligned}
\label{eq:portCapOptdual}
\end{equation}
This problem is convex in the variable $\nu$. The tightest bound on the solution of the original problem is found when the dual problem is minimized over $\nu$~\cite{Boyd+Vandenberghe2004}. The solution to this problem is found in~\cite{Ehrenborg+Gustafsson2020} by incorporating the matrices of the first condition in~\eqref{eq:portCapOptdual} into the channel matrix of the system, see Appendix~\ref{app:singval}. This recasts the system in a form where it can be solved by water-filling~\cite{Paulraj+etal2003,Molisch2011}. 

The water-filling procedure is carried out by taking the singular value decomposition of the channel matrix. Each singular value represents the loss associated with feeding power in its corresponding mode. The optimal solution is found by iteratively allocating energy to the channels associated with the largest singular values~\cite{Paulraj+etal2003,Molisch2011,Ehrenborg+Gustafsson2020}. The algorithm can be expressed as,
\begin{equation}
\begin{aligned}
	& \maximize && \sum_{n=1}^{N} \log_2\left(1 + a_n\gamma\sigman^2 \right)\\
	& \subto && \sum_{n=1}^{N} a_n = 1, 
\end{aligned}
\label{eq:waterfilling}
\end{equation}
where $a_n\geq 0$ is the power allocation fraction in each channel, and $\sigman$ is the singular value of the corresponding channel. 

The singular values of the channel matrix are
\begin{equation}
\sigman^2=\dfrac{1 + \nu}{\emod/ Q+\nu},
\label{eq:SVD_channel}
\end{equation}
where $\emod$ are the eigenvalues of a set of modes, here referred to as energy modes~\cite{Gustafsson+etal2016a}. The energy modes are calculated through the generalized eigenvalue problem, defined here over port mode matrices,
\begin{equation}
\mat{w}_\mathrm{x}\portv_n = \emod \mat{r}\portv_n,
\label{eq:eig_Emodes_port}
\end{equation}
where~$\portv_n$ are the modal port voltages. These modes are similar to characteristic modes~\cite{Harrington+Mautz1971} in the sense that they have the property of orthogonal radiation patterns. In addition, they minimize the total energy stored by the antenna. This has the effect of implicitly maximizing the bandwidth of the modes with the lowest eigenvalues. 

\begin{figure}
    \centering
    \includegraphics[width=\linewidth]{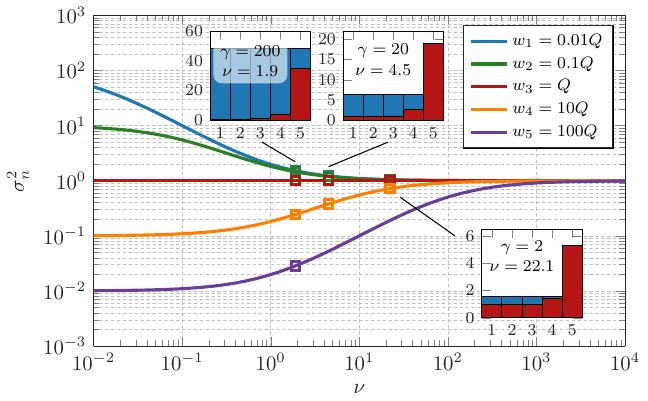}
    \caption{The singular values~\eqref{eq:SVD_channel} as a function of $\nu$ for different energy mode values. The three insets show water-filling solutions~\eqref{eq:portCapOptdual} at optimal $\nu\in\{22.1,4.5,1.9\}$ for three different \ac{SNR} values, $\gamma\in\{2,20,200\}$. For $\gamma\in\{2,20\}$ the weakest mode, $w_5$, is not used because the \ac{SNR} value is insufficient.}
    \label{fig:singularvaluesQ}
\end{figure}

The singular values $\sigma^2_n$ from~\eqref{eq:SVD_channel} are depicted in Fig.~\ref{fig:singularvaluesQ} as a function of $\nu$ for energy modes with eigenvalues chosen as $w_n/Q=10^{n-3}$, where $n\in\{1,\dots,5\}$. The water-filling solutions~\eqref{eq:waterfilling} of~\eqref{eq:portCapOptdual} at the optimal $\nu$ are shown in insets for the SNR values~$\gamma\in\{2,20,200\}$. We observe that the optimal $\nu$ decreases with increasing~$\gamma$ and that strong channels, corresponding to $w_n/Q\leq 1$, are included for all considered~$\gamma$, whereas the weaker channels, $w_n/Q>1$, are only included for high~$\gamma$. The singular values approach unity as $\nu\to\infty$ with the explicit equal power allocation solution for $N$ equal channels
\begin{equation}
    C = N\log_2\left(1+\frac{\gamma}{N}\right).
    \label{eq:CapIdealN}
\end{equation}
This is also the explicit solution for the case with $\max(w_n/Q)\leq 1$, because the Q-factor constraint in~\eqref{eq:MIMOoptQport} is trivially satisfied.
This suggests that the number of available channels for a given Q-factor can be estimated by studying the energy mode eigenvalues. Modes with $w_n/Q\leq 1$ can be considered available and will always be used by the system to induce maximum spectral efficiency. The energy modes not only simplify the optimization problem for maximum capacity to water-filling solutions~\eqref{eq:waterfilling} but also provide a physical interpretation of the results. Let us now consider a few examples to see how the solution to this optimization problem behaves.

\subsection{Example -- Two Dipoles of Variable Separation Distance}
\label{sec:2dip_vardist}

The first example deals with two thin-strip dipoles of (resonant) length~$L/\lambda = 0.49$ and width~$L/W = 50$. The dipoles are made of perfect electric conductor (PEC) and separated by the distance~$d$ swept in the interval~$d/L \in [0, 2]$. 

\begin{figure}
\includegraphics[width=\columnwidth]{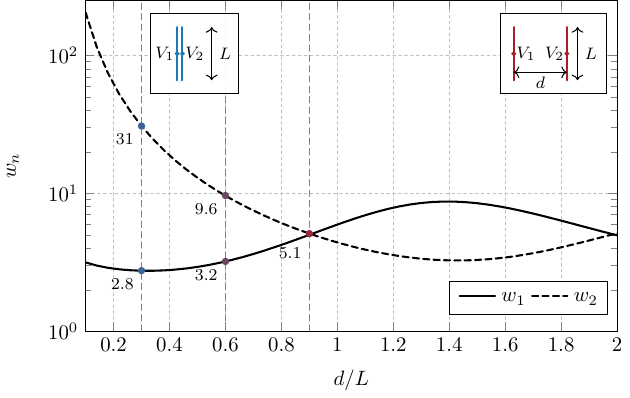}
\caption{The eigenvalues of the energy port-modes~$w_n$ evaluated by~\eqref{eq:eig_Emodes_port} for an array of two dipoles placed in-parallel, separated by the distance~$d$ (see the insets). The values of~$w_n$ are marked for separation distances~$d/L\in \{0.3, 0.6, 0.9 \}$.}
\label{fig:portModes2dipModes}
\end{figure}

\begin{figure}
    \centering
    \includegraphics[width=\columnwidth]{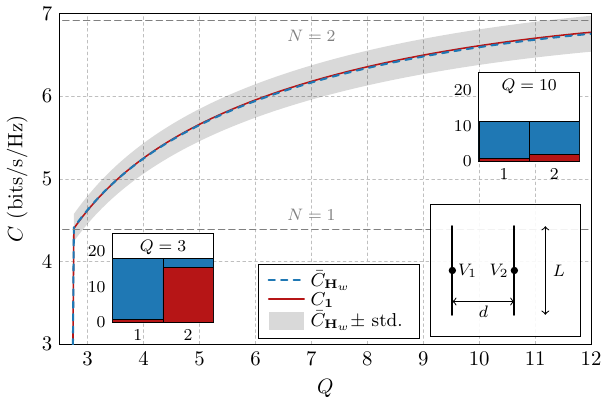}
    \caption{The maximal ergodic spectral efficiency of two centrally fed dipoles in Rayleigh fading (the blue dashed line) and deterministic isotropic (the solid red line) channels. The gray area denotes the one standard deviation limit of the channel realizations. The dipoles have electrical size $L/\lambda=0.49$, are separated $d/L=0.3$, and the \ac{SNR} has been set to $\gamma=20$. The dashed horizontal lines correspond to one and two ideal equal power allocation channels~\eqref{eq:CapIdealN}. The two insets show the water-filling solutions for $Q=3$ and $Q = 10$.}
    \label{fig:2dip_rayleigh}
\end{figure}

The sole input into the optimization procedure is the set of energy mode eigenvalues~\eqref{eq:eig_Emodes_port}. Their dependence on the separation distance~$d/L$ is shown in Fig.~\ref{fig:portModes2dipModes}. There are two types of modes, in-phase and out-of-phase modes, which become degenerated at $d/L\approx 0.9$ and $d/L \approx 1.9$. The smallest eigenvalue of the in-phase mode is $w_1 \approx 3$ for small distances whereas the out-of-phase mode $w_2$ increases rapidly as $d/L\to 0$. This implies an onset around $Q\approx 3$ with a spectral efficiency of a single channel, $N=1$, used in~\eqref{eq:CapIdealN}. The out-of-phase mode starts to contribute for larger Q-factors as $w_n/Q$ decreases in the channel singular values~\eqref{eq:SVD_channel}, see also Fig.~\ref{fig:singularvaluesQ} and the insets in Fig.~\ref{fig:2dip_rayleigh} with the water-filling solutions for $Q = 3$ and $Q = 10$. At the degeneracy around $d/L\approx 0.9$ the modes contribute equally following the spectral efficiency~\eqref{eq:CapIdealN} from two equal channels ($N=2$) as $Q$ increases. For larger distances, the two modes have similar eigenvalues~$w_n$ as the mutual coupling diminishes.

The maximum spectral efficiency calculated with~\eqref{eq:MIMOoptQport} is depicted in Fig.~\ref{fig:2dip_rayleigh} for $\gamma=20$ and spacing $d/L=0.3$ as a function of the Q-factor. The onset is bounded by the lower bound on the Q-factor, determined from the smallest eigenvalue~$\emod$. For two dipoles separated by distance $d/L \approx 0.3$ that bound is just below $Q = 3$ as seen in Fig.~\ref{fig:portModes2dipModes}. This supports previous observations~\cite{Hazdra+etal2011,Munk2003} that two mutually coupled dipoles might have lower $Q$ than a single dipole (here $Q \approx 5$).
After the onset the  spectral efficiency $C$ increases towards $C=6.9$ which is reached for $Q\approx 31$. The horizontal dashed lines in Fig.~\ref{fig:2dip_rayleigh} show the spectral efficiency of one and two ideal equal power allocation channels that do not take into account antenna geometry or channel effects. These lines serve as an indicator of how many pathways the system is utilizing~\cite{Ehrenborg+Gustafsson2020}. The Q-factor values at which the two dipoles produce the optimal spectral efficiency of either one or two ideal channels correspond perfectly to the first and second eigenvalue from Fig.~\ref{fig:portModes2dipModes}. The two dipoles can be seen to have solutions in between one and two ideal channels. Indicating that they are utilizing two channel pathways.

The optimization problem~\eqref{eq:portCapOptdual} can be solved for a \ac{MIMO} antenna in a complex propagation channel, by multiplying the matrix~$\mat{s}$ with the channel matrix. However, the \ac{svd} of the channel matrix~\eqref{eq:SVD_channel} can no longer be found efficiently. For each~$\nu$ considered in the optimization process, the \ac{svd} must be calculated numerically. In Fig.~\ref{fig:2dip_rayleigh}, the maximum spectral efficiency in a uniform Rayleigh fading channel and a deterministic isotropic channel are compared for the two dipoles. It can be seen that the antenna in the deterministic channel follows the behaviour of the system in a Rayleigh channel. This shows that the deterministic channel can be used to estimate the maximum spectral efficiency characteristics of an electrically small antenna design problem in a uniform Rayleigh fading multipath environment.

In Fig.~\ref{fig:2dip_rayleigh}, the spectral efficiency grows with~$Q$. This can be misleading, and does not mean that higher capacity can be achieved by a narrow-band system. Spectral efficiency is a measure of the maximum bit rate that can be achieved on a $1\unit{Hz}$ bandwidth. By allowing a higher Q-factor in the optimization problem the solution can concentrate all spectral efficiency to a specific frequency. The true capacity can be calculated from the spectral efficiency by multiplying it by the bandwidth. An indication of the capacity can be found by considering the spectral efficiency divided by the required Q-factor,~$C/Q$. This measure is adopted for the remaining results in the paper.

\begin{figure}
\includegraphics[width=\columnwidth]{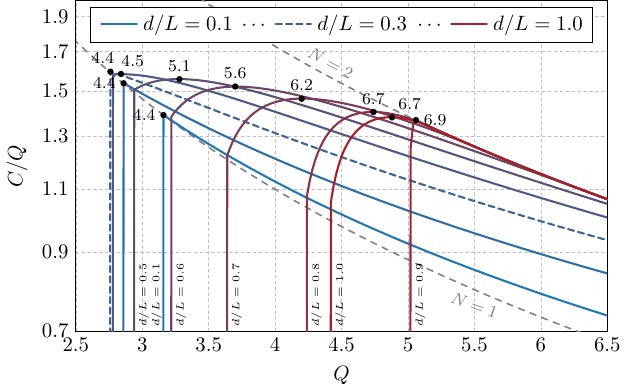}%
\caption{The maximal spectral efficiency over Q for two thin-strip dipoles of separation distance~$d$.  Dot marks with numbers correspond to particular spectral efficiencies, evaluated at $C/Q$ maximum. The \ac{SNR} has been set to $\gamma=20$.}
\label{fig:portModes2dipDistance}
\end{figure}

The spectral efficiency~$C$ over $Q$, evaluated via~\eqref{eq:MIMOoptQport}, for two dipoles as a function of their separation distance $d/L$ is shown in Fig.~\ref{fig:portModes2dipDistance}.  It is seen that there exists an optimal distance, $d/L \approx 0.3$, for which $C/Q$ is maximal. This is the distance with the lowest possible eigenvalue in Fig.~\ref{fig:portModes2dipModes}. All the curves are bounded from the left by the fundamental bound on $Q$ for two dipoles given by the lowest eigenvalue $w_n$. The maximum spectral efficiency is bounded by~\eqref{eq:CapIdealN} with $N=2$, shown by a dashed curve, and is first reached for the degenerate case around the distance~$d/L\approx 0.9$ with $w_1=w_2\approx 5.1$, see Fig.~\ref{fig:portModes2dipModes}. However, this curve does not dominate the others, even though it has the highest spectral efficiency. This is due to higher~$Q$ of the dipoles. For distances between $d/L = 0.3$ and $d/L = 0.9$, we see a gradual increase from the lowest $Q=w_1$ with a $N=1$ spectral efficiency~\eqref{eq:CapIdealN} increasing to two channels $N=2$ as $Q$ approaches $w_2$.
 
When the dipoles are separated by a large distance they act as separate radiators and do not improve their joint Q-factor. This indicates that there is a trade-off between decoupling the dipole radiators to improve spectral efficiency and utilizing their mutual coupling to improve their bandwidth~\cite{Hannula+etal2016}.

\subsection{Example -- Two Dipoles Rotated by An Arbitrary Angle}
\label{sec:2dip_angle}
\begin{figure}
\includegraphics[width=\columnwidth]{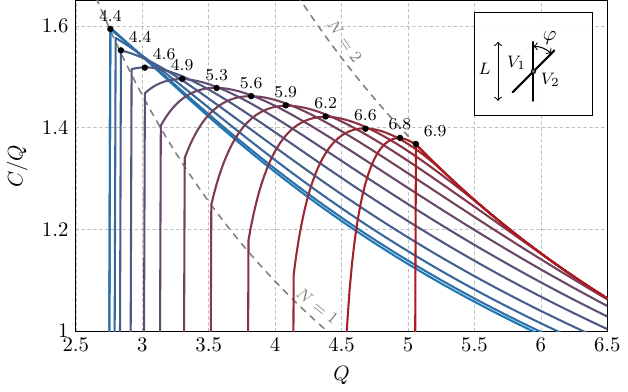}
\caption{The maximal spectral efficiency over $Q$ for two thin-strip dipoles rotated by the angle~$\varphi$. The dot marks with numbers correspond to particular spectral efficiencies, evaluated at $C/Q$ maximum. The \ac{SNR} has been set to $\gamma=20$.}
\label{fig:portModes2dipAngle}
\end{figure}

Here, we study the dependence on the angle between the dipoles. The separation distance between two dipoles is fixed to the optimal value from the previous example, \ie $d/L = 0.3$. The dipoles are of the same dimensions, located at $x=-d/2$ and $x=d/2$, and the left dipole is parallel to the $z$-axis. The axis of rotation for the right dipole coincides with the $x$-axis, see the insets of Fig.~\ref{fig:portModes2dipAngle}.

\begin{figure}
\includegraphics[width=\columnwidth]{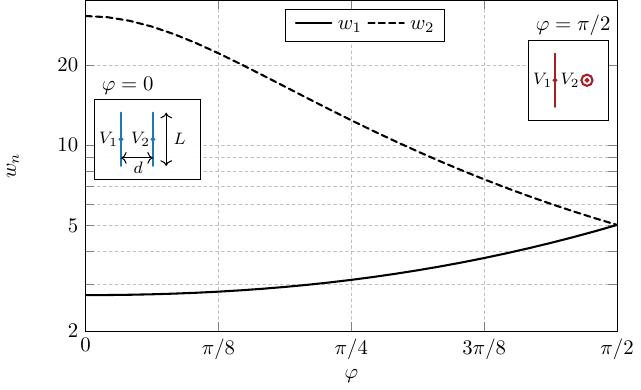}
\caption{The eigenvalues of energy port-modes~$w_n$ evaluated by~\eqref{eq:eig_Emodes_port} for an array of two dipoles placed in-parallel and rotated by angle~$\varphi$.}
\label{fig:portModes2dipAngleModes}
\end{figure}

The input to the optimization are the stored energy eigenvalues~$w_n$, depicted in Fig.~\ref{fig:portModes2dipAngleModes}. The eigenvalues have values $w_1 = 2.8$ and $w_2 = 31$ for the rotation angle~$\varphi = 0$. This agrees with the values in Fig.~\ref{fig:portModes2dipModes} for $d/L=0.3$ since the arrangements are the same, and a doublet of in-phase and out-of-phase modes exist. The eigenvalues become degenerate for angle~$\varphi=\pi/2$, \ie both states can be utilized equally with expected onset around $Q \approx 5$.

The results of~$C/Q$ are depicted in Fig.~\ref{fig:portModes2dipAngle}. The curve for~$\varphi = 0$ is identical to blue dashed one in Fig.~\ref{fig:portModes2dipDistance} for~$d/L$ with the same onset at~$Q \approx 2.8$. Similar trends as in Fig.~\ref{fig:portModes2dipDistance} are observed. The spectral efficiency over $Q$ decreases with an increasing angle of rotation~$\varphi$. Conversely, the spectral efficiency increases and reaches its maximum for $\varphi=\pi/2$, where the dipoles are crossed and able to utilize two times more spherical harmonics, \ie their radiation is relatively uncoupled, and, as confirmed by the agreement with the asymptote for $N=2$ (the gray dashed line), both stored energy modes are exploited. This is confirmed by the identical value of spectral efficiency, $C \approx 6.9$ for two distant dipoles in Fig.~\ref{fig:portModes2dipDistance} and two dipoles rotated by angle~$\varphi=\pi/2$ in Fig.~\ref{fig:portModes2dipAngle}. Similarly, as in the previous example, the cost for this performance in capacity is higher required bandwidth (lower~$Q$). 

\subsection{Example -- A Dipole Array of Various Number of Elements}
\label{sec:dip_array}
The last example of this section studies the optimal capacity depending on the increasing number of dipoles in the linear array. Three separation distances, $d/L \in \{ 0.3, 0.5, 0.7 \}$, are studied, respectively. The dimensions of the individual radiators are the same as in the previous examples.

\begin{figure}
\includegraphics[width=\columnwidth]{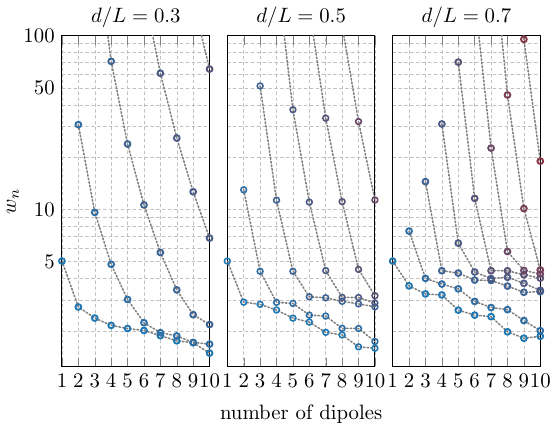}
\caption{The eigenvalues of stored energy modes~$w_n$ for a dipole array consisting of a variable number of thin-strip dipole elements (from one to ten). The spacing between dipoles is $d/L \in \{ 0.3, 0.5, 0.7\}$.}
\label{fig:portModes1to10dipolesModes}
\end{figure}

The stored energy eigenvalues~$w_n$ are depicted in Fig.~\ref{fig:portModes1to10dipolesModes} in dependence on number of dipoles and separation distance~$d/L$. The eigenvalues monotonically decrease with adding more dipoles into an array. This effect is emphasized with increasing separation distance~$d/L$, \ie{} more modes are closer to the dominant eigenvalue~$w_1$ for $d/L = 0.7$ than for $d/L = 0.3$ and $d/L = 0.5$. This has immediate effect on the results of the optimization.

\begin{figure}
\includegraphics[width=\columnwidth]{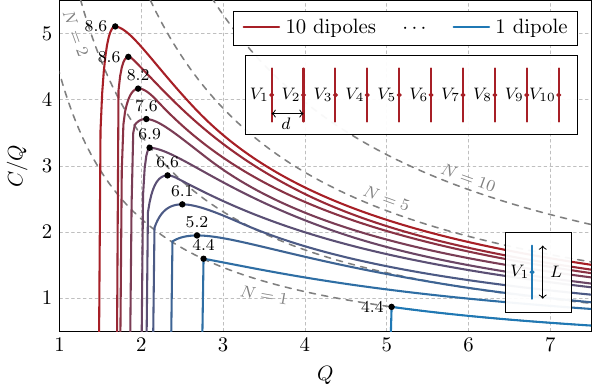}
\caption{The optimized~$C/Q$ and the spectral efficiency~$C$ for the dipole array. The spacing between the dipoles is $d/L = 0.3$, \ie{} $kd = 0.92n$, $n \in \{1,\dots,10\}$. The dot marks with the numbers correspond to particular spectral efficiencies, evaluated at $C/Q$ maximum. The \ac{SNR} has been set to $\gamma=20$.}
\label{fig:portModes1to10dipoles}
\end{figure}

\begin{figure}
\centering
\includegraphics[width=\columnwidth]{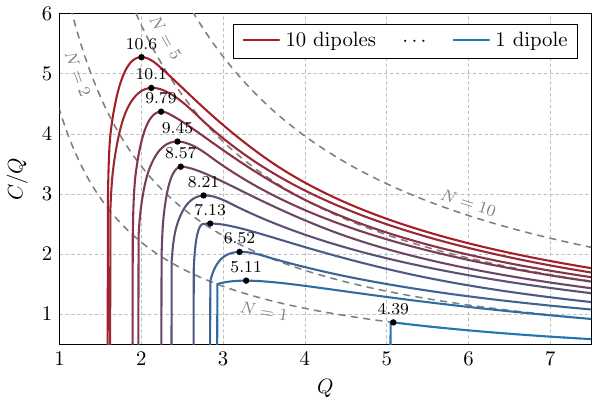}
\caption{The optimized~$C/Q$ and the spectral efficiency~$C$ for the dipole array. The spacing between the dipoles is $d/L = 0.5$, \ie{} $kd = 1.54n$, $n \in \{1,\dots,10\}$. The dot marks with the numbers correspond to particular spectral efficiencies, evaluated at $C/Q$ maximum. The \ac{SNR} has been set to $\gamma=20$.}
\label{fig:portModes1to10dipolesB}  
\end{figure}

\begin{figure}
\centering
\includegraphics[width=\columnwidth]{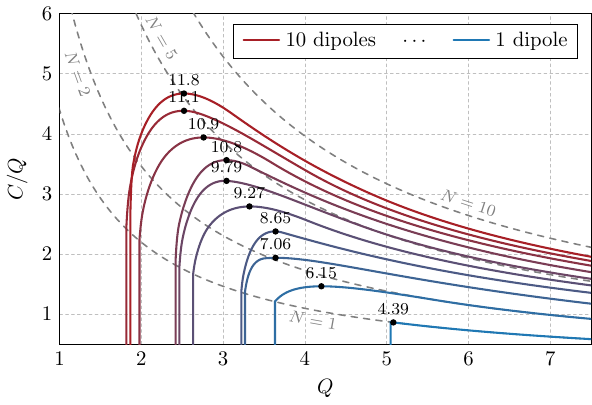}
\caption{The optimized~$C/Q$ and the spectral efficiency~$C$ for the dipole array. The spacing between the dipoles is $d/L = 0.7$, \ie{} $kd = 2.16n$, $n \in \{1,\dots,10\}$. The dot marks with the numbers correspond to particular spectral efficiencies, evaluated at $C/Q$ maximum. The \ac{SNR} has been set to $\gamma=20$.}
\label{fig:portModes1to10dipolesC}
\end{figure}

The change in spectral efficiency constrained by~$Q$ is depicted in Figs.~\ref{fig:portModes1to10dipoles}--\ref{fig:portModes1to10dipolesC}. It is obvious that the performance is significantly boosted with each dipole added to the array and, for a given number of radiators, the maximum of $C/Q$ and the maximum of $C$ coincide. The cost of the higher spectral efficiency is an increase of electrical size. Adding more dipoles imply more available states for the optimization. The high-order eigenvalues grow rapidly, whereas the other ones are accumulated closer to the dominant eigenvalue~$w_1$, introducing additional degrees of freedom into the radiation process, see Fig.~\ref{fig:portModes1to10dipolesModes}. This is most apparent for~$d/L = 0.7$, Fig~\ref{fig:portModes1to10dipolesC}, where the $C/Q$ for $10$ dipoles approaches the asymptote for~$N=10$ (gray dashed line) even for relatively low values of Q. This is because of large number of stored energy modes have comparably low values, $w_n < 5$, when the number of dipoles reach~$10$, see Fig.~\ref{fig:portModes1to10dipolesModes}, right pane.

\section{Bound on Antenna Design}
\label{sec:currents}
In Section~\ref{sec:ports} the spectral efficiency is optimized through input parameters to ports for a fixed geometry. In order to formulate an upper bound on spectral efficiency that can be gained through antenna design we must make the antenna structure part of the optimization variable. In this section this is done through current optimization~\cite{Gustafsson+etal2016a}. 

An antenna structure, \eg the structure depicted in Fig.~\ref{fig:portMIMO}b, radiates electromagnetic fields through the current that the excitation in the ports induces across it. In \ac{MoM} this current is modelled through equivalent currents in free space. The equivalent current is represented through the basis functions that are connected to the mesh. Consider a \ac{MoM} mesh covering the entire design region in Fig.~\ref{fig:portMIMO}a as in Fig.~\ref{fig:portMIMO}c, if this mesh is sufficiently dense it can support the equivalent current of any antenna structure that could be designed in the region. The technique of current optimization optimizes the weights of the basis functions of such a mesh. Because the equivalent current can represent the radiation properties of any antenna in the design region, an optimal current bounds those properties. However, antenna designs have several physical limitations to the current that they can induce, such as feeding. The equivalent current does not share these restrictions and therefore can support current distributions that are not realizable but perform better than any antenna design. This ensures that the optimal value gained through current optimization is always equal to or greater than that gained through antenna design in the same region. 

 Optimizing spectral efficiency through current optimization has been treated extensively by the authors in~\cite{Ehrenborg+Gustafsson2018,Ehrenborg+Gustafsson2020}, and is only briefly reiterated here. 

The optimization problem~\eqref{eq:MIMOoptQport} can readily be rewritten in terms of the equivalent current. Take the covariance of the current, $\Pm=\medel{\Jm\Jm^\herm}/2$, as the optimization variable and replace all port mode matrices by their equivalent \ac{MoM} matrices, see~\eqref{eq:portMode4}. The optimization problem is then written as
\begin{equation}
\begin{aligned}
	& \maximize && \log_2\det\left(\Id+\gamma\Sm\Pm\Sm^{\herm}\right) \\
	& \subto && \Tr(\mat{W}_\mrm{x}\Pm) \leq Q &&& \\
	&  &&  \Tr(\Rm\Pm) = 1\\
	& && \rank\Pm \leq N  \\
    & && \ \Pm \succeq 0,
\end{aligned}
\label{eq:convexMIMOcurr}	
\end{equation}
where $N$ is the number of degrees of freedom of the system. Here, the port positions are no longer fixed, only their number is specified. This problem can be solved in different ways depending on what we wish to investigate.

The optimization problem~\eqref{eq:convexMIMOcurr} can be made convex by dropping the rank constraint~\cite{Ehrenborg+Gustafsson2020}. The problem can then be solved directly using freely available convex optimization software such as CVX~\cite{Grant+Boyd2011,Ehrenborg+Gustafsson2018}. However, this method can become computationally cumbersome for bigger problems. 

The problem~\eqref{eq:convexMIMOcurr} can instead be solved using the method detailed in~\cite{Ehrenborg+Gustafsson2020}, see Appendix~\ref{app:singval}. The solution is equivalent to replacing the singular values in~\eqref{eq:waterfilling}, with those calculated through~\eqref{eq:SVD_channel} using the $N$ lowest energy mode eigenvalues~$W_n$ calculated by the generalized eigenvalue problem,
\begin{equation}
\mat{W}_\mrm{x} \Jm_n = W_n\Rm\Jm_n.
\label{eq:eig_Emodes_curr}
\end{equation}
These modes have the same properties as their port equivalents~\eqref{eq:eig_Emodes_port}, \ie orthogonal radiation patterns and minimization of the total stored energy. 

\subsection{Modal Analysis for a rectangular plate}
\label{sec:curr_rayleigh}

\begin{figure}
    \centering
    \includegraphics[width=\columnwidth]{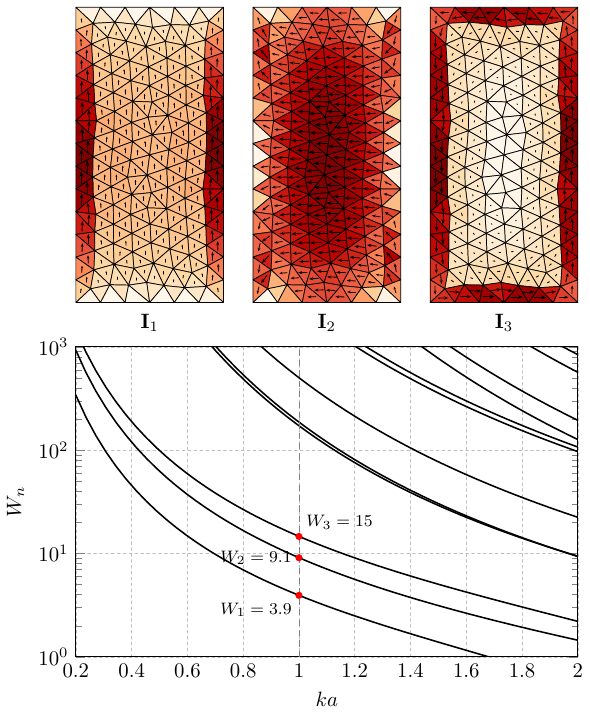}
    \caption{Energy mode eigenvalues $W_n$~\eqref{eq:eig_Emodes_curr} for a rectangular region with side lengths $\ell\times\ell/2$, the first three eigencurrents at $ka = 1$ are shown as well.}
    \label{fig:plate_mod}
\end{figure}

\begin{figure}
\centering
\includegraphics[width=\columnwidth]{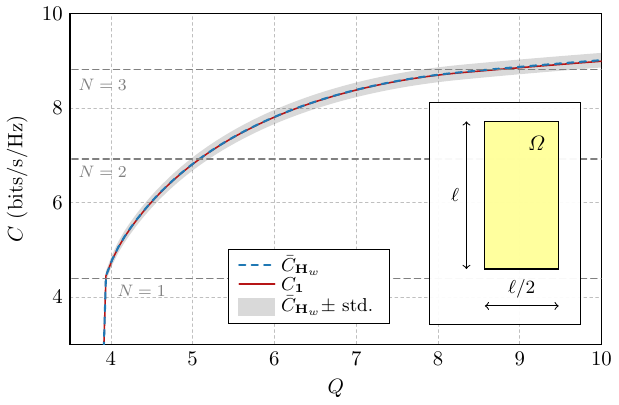}
\caption{Maximum ergodic spectral efficiency of a $ka=1$, $\ell\times\ell/2$ plate in a uniform Rayleigh fading channel (blue dashed line) and a deterministic isotropic channel (solid red line). The gray area denotes one standard deviation from the average of the Rayleigh channel case. The dashed horizontal lines correspond to one, two, and three ideal equal power allocation channels~\cite{Ehrenborg+Gustafsson2020}.}
\label{fig:CapvQ_rayleigh}
\end{figure}

Fig.~\ref{fig:plate_mod} depicts the eigenvalues of energy modes for a rectangular plate with side lengths $\ell\times\ell/2$ as a function of electrical size $0.2\leq ka\leq 2$. The modal behaviour of two electric dipoles and one magnetic dipole dominate for electrically small structures, where the scaling $W_n\sim (ka)^{-3}$ is found~\cite{Chu1948,Gustafsson+etal2007a}. Higher-order modes have very large eigenvalues $W_n$ and require similarly high $Q$ to contribute. The energy-mode eigenvalues $W_n$ decrease with increasing electrical size and several modes are available for larger $ka$. For $ka=1$, we have the first eigenvalue~$W_1\approx 4$, proposing an onset around $Q=4$, and two additional eigenvalues at $9$ and $15$  indicate three independent channels below $Q=15$, see Fig.~\ref{fig:CapvQ_rayleigh}.
 
When optimizing over the current density the number of channels in~\eqref{eq:capacity} can no longer be considered small. Therefore, we must analyze the impact of the Rayleigh channel on the optimization results. Consider~\eqref{eq:convexMIMOcurr} with the objective function,
\begin{equation}
    C_{\mat{H}_\mrm{w}}= \log_2\det\left(\Id+\gamma\Hmw\Sm\Pm\Sm^{\herm}\Hmw^{\herm}\right) .
\end{equation}
This can be done by numerically calculating the SVD of the channel matrix for each~$\nu$.

In Fig.~\ref{fig:CapvQ_rayleigh} the current on a $ka=1$ plate has been optimized for maximal spectral efficiency with and without the inclusion of a uniform Rayleigh channel. The average of the uniform Rayleigh fading channel realizations has the same performance as the plate in the deterministic isotropic spherical channel. This supports the claims in~\cite{Ehrenborg+Gustafsson2018,Ehrenborg+Gustafsson2020} that the Rayleigh channel can be neglected when studying the fundamentals of antenna design for optimal spectral efficiency in uniform multipath environments. 

The standard deviation of the Rayleigh channel realizations for the plate in Fig.~\ref{fig:CapvQ_rayleigh} is significantly smaller than what can be seen for two dipoles in Fig.~\ref{fig:2dip_rayleigh}. This is due to the added degrees of freedom in optimizing an entire plate. The plate can support higher order modes at lower Q-values and its current can be tailored to specific channel realizations. 

\subsection{Sub-region Energy modes}
\label{sec:curr_subreg}

Antennas inside communication devices are, in general, much smaller than the total device size~\cite{Wong2003}. This means that only a sub-region of the device is dedicated to antenna design. The antennas therein excite currents across the entire device that contribute to communication. From an optimization perspective this can be seen as only controlling the current within the antenna sub-region. The optimization of arrays through their port quantities in Section~\ref{sec:ports} is similar to this concept. The ports control the current on the dipole on a very small sub-region of the antenna.  In this section this concept is generalized to controlling the current in larger sub-regions, see Fig.~\ref{fig:portMIMO}c. Optimizing the current on a sub-region for maximal spectral efficiency calculates a bound on the spectral efficiency available from designing an antenna in a sub-region of a device. This is interesting to investigate since bandwidth and Q-factor are usually harshly restricted by reducing the antenna size~\cite{Volakis+etal2010,Capek+etal2017b}. 


\begin{figure}%
\includegraphics[width=\columnwidth]{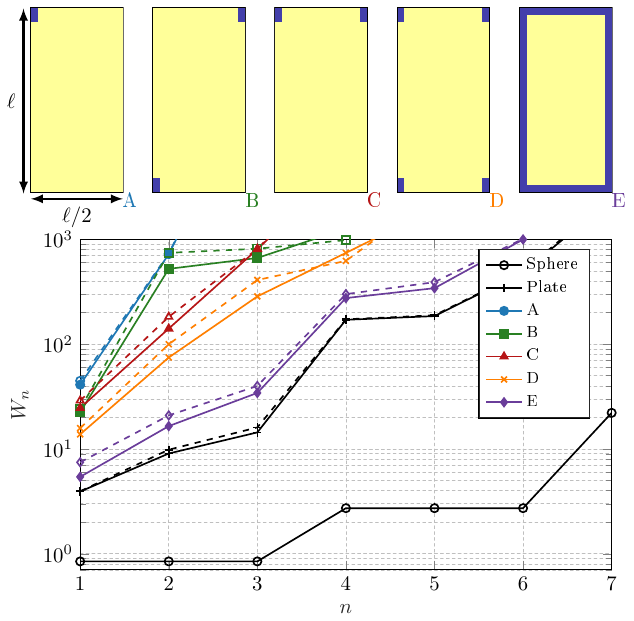}%
\caption{The eigenvalues~$W_n$ of a spherical shell, a rectangular plate with aspect ratio $2:1$, and several of its sub-region arrangements with size $0.1\ell\times0.05\ell$ for the cases A-D and width $0.05\ell$ for the case E, where the current is controlled inside the blue regions. The electrical size is $ka=1$, where $k$ is the wave number, and $a$ is the radius of the smallest sphere circumscribing the entire structure. The dashed lines correspond to the characteristic mode currents evaluated in the Rayleigh quotient~\eqref{eq:Rayleigh}~\cite{Alroughani+etal2016}.}%
\label{fig:subModeStr}%
\end{figure}

Consider the placement of sub-regions for antenna design within a device. Which orientation of sub-region has the greatest potential spectral efficiency can be evaluated by comparing their energy mode eigenvalues. In Fig.~\ref{fig:subModeStr}, five different sub-region eigenvalues, as well as the plate and circumscribing spherical shell eigenvalues, are shown. By studying how close each of the sub-region cases come to the eigenvalues of the full plate we can evaluate how well they induce the full performance of the plate. In Fig.~\ref{fig:subModeStr}, we see that there is a gap between the eigenvalue of the full plate compared to small segregated sub-regions, whereas the case~E, the loop region, has eigenvalues close to the full plate. We see that the two diagonally placed elements in case~B outperform the two elements in case~C for the first and third mode. This is similar to the case in~\cite{Ehrenborg+Gustafsson2020} due to the first and second order modes being induced diagonally across the plate. Therefore, the two diagonally situated sub-regions do not effectively induce the second mode across the opposite diagonal. The eigenvalues of the circumscribing spherical shell are significantly lower than those of the plate. These modes are grouped in sets of three because the spherical shell supports three degenerate dipole modes.

The dashed lines in Fig.~\ref{fig:subModeStr} correspond to the characteristic mode currents evaluated in the Rayleigh quotient of the eigenvalue problem~\eqref{eq:eig_Emodes_curr}, \ie
\begin{equation}
    W_{\mrm{c},n} = \frac{\Jm_{\mrm{c},n}^{\herm}\mat{W}_\mrm{x}\Jm_{\mrm{c},n}}{\Jm_{\mrm{c},n}^{\herm}\Rm\Jm_{\mrm{c},n}} ,
    \label{eq:Rayleigh}
\end{equation}
where $\Jm_{\mrm{c},n}$ are the characteristic mode currents~\cite{Harrington+Mautz1971}. These values indicate how well the characteristic mode currents perform in the metric of energy modes. In Fig.~\ref{fig:subModeStr}, we see that the characteristic modes basically overlap the energy mode eigenvalues for all considered cases. This indicates that the characteristic modes are a good tool for \ac{MIMO} antenna design, validating previous design methodologies~\cite{Li+etal2014,Miers+etal2013,Manteuffel+Martens2011}.

\subsection{Sub-region Optimization}

\begin{figure}
    \centering
    \includegraphics[width=\linewidth]{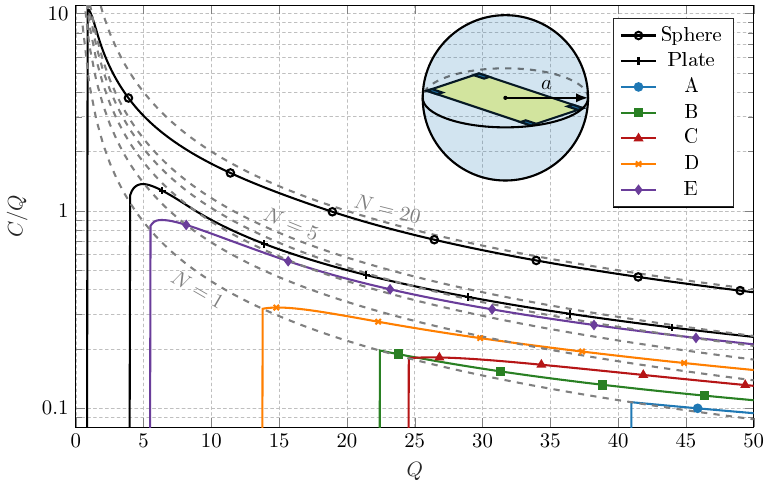}
    \caption{The optimal spectral efficiency over $Q$ for the $ka=1$ spherical shell, plate with aspect ratio $2:1$, and its sub-regions presented in Fig.~\ref{fig:subModeStr}, for different Q-factor restrictions. The dashed gray lines correspond to increasing numbers of ideal equal power allocation channels. The \ac{SNR} has been set to $\gamma=20$.}
    \label{fig:subCapvsQ}
\end{figure}

In Fig.~\ref{fig:subCapvsQ}, the spectral efficiency over $Q$ is depicted for the electrical size $ka=1$ as a function of the required Q-factor using the energy modes depicted in Fig.~\ref{fig:subModeStr}. This Pareto-type curve delimits the feasible region of the problem and reveals that the capacity and the Q-factor are strictly conflicting parameters. It can be seen that there is a preferred Q-factor for optimal spectral efficiency for different sizes and geometries found at the maximum of the curves. This maximum is found close to the minimum feasible Q-factor found in Fig.~\ref{fig:subCapvsQ}. However, the performance in terms of $C/Q$ deteriorates slowly with increasing Q-factor. This indicates that sub-region \ac{MIMO} antennas use less bandwidth but more spatial diversity to maintain high performance.

The dashed gray lines in Fig.~\ref{fig:subCapvsQ} depict the equal power allocation channels~\eqref{eq:CapIdealN}. When the solution of the different sub-region cases passes one of these lines it is utilizing at least that many different channels to induce optimal spectral efficiency. This number is the structure's number of effective modes for a certain allowed Q-factor.
The number of effective modes can be estimated from Fig.~\ref{fig:subModeStr} by counting the number of eigenvalues below $W_n=Q$. This is the limit for what is considered an effective mode, see Fig.~\ref{fig:singularvaluesQ}. When the number of ports is restricted by a certain number, in~\eqref{eq:convexMIMOcurr}, the spectral efficiency over $Q$ maximally reaches the dashed gray curve associated with that number of ideal channels. 

It is evident that the small segregated sub-regions from the case A to D do not effectively induce the available spectral efficiency of the full plate, as predicted from Fig.~\ref{fig:subModeStr}. However, the loop region reaches $C/Q$ values much closer to the full plate. In contrast to the case studied in~\cite{Ehrenborg+Gustafsson2020}, that was restricted by efficiency, this gap between the sub-regions and the full plate remains the same when the size of the plate is reduced. Instead of narrowing the gap between the full plate and its segregated sub-regions, as in~\cite{Ehrenborg+Gustafsson2020}, a reduction of size makes the sub-region solution unfeasible, due to the lower Q-factor bound~\cite{Cismasu+Gustafsson2014a}. Therefore, we can conclude that the optimal performance of the embedded \ac{MIMO} antennas is more restricted by the limited bandwidth than the requirements on radiation efficiency.

\section{Conclusions}
\label{sec:conclusions}

It has been shown that a set of modes known as energy modes, as well as characteristic modes, serves as a useful design tool when analyzing performance of \ac{MIMO} antennas. The conclusion that a set of properly weighted characteristic modes approximately form an optimal current for capacity can be utilized in practice, specifically since simultaneous excitation of characteristic modes is well explored. However, due to the harsh penalties on stored energy when reducing the design region, it has been concluded that it is not possible to reach the full potential of the plate while only feeding it with a set of small segregated sub-regions. However, the use of a connected loop region was shown to be much more efficient.

Spectral efficiency bounds of \ac{MIMO} antennas in uniform Rayleigh fading channels and a deterministic isotropic channel have been considered when restricted by the Q-factor. It was shown that the uniform Rayleigh fading channel does not affect the characteristics of the spectral efficiency bound. The Pareto-type bound has been illustrated in spectral efficiency over~$Q$, where a certain Q-factor was shown to be Pareto optimal. 

The modal analysis illustrated in this paper could also be carried out on designed antennas to evaluate their adherence to the principals suggested here. Generalizing this method to include several different design parameters remains an interesting future prospect. 

\appendices

\section{Rayleigh Channel Contribution}
\label{app:ray}

To study the dependence of the system on the uniform Rayleigh fading channel, consider a \ac{MIMO} system with a finite and small number of ports $N$. Since the number of ports is small, the \ac{SNR} can be considered to be large in each port, \ie $\gamma\Hmw\Sm_{\mrm{2,1}}\Am\Sm_{\mrm{2,1}}^{\herm}\Hmw^{\herm} \succ \alpha \Id$, $\alpha \gg 1$. The ergodic spectral efficiency is then gained by averaging~\eqref{eq:capacity} over channel realizations,
\begin{equation}
\begin{split}
    \bar{C} = \, & \medel{\log_2\det(\gamma\Hmw\Sm_{\mrm{2,1}}\Am\Sm_{\mrm{2,1}}^{\herm}\Hmw^{\herm})}   \\ 
   = \, &\log_2\det(\gamma) + \medel{\log_2\det(\Hmw)} \\ 
    &+\medel{\log_2\det(\Hmw^\herm)} + \log_2\det(\Sm_{\mrm{2,1}}\Am\Sm_{\mrm{2,1}}^{\herm}). 
\end{split}    
\label{eq:Ergodic_cap}
\end{equation}
The effect of the channel in such a system is only an additive term in the spectral efficiency evaluation.  Consider the correlation loss, defined as 
\begin{equation}
    \Delta \bar{C} = \bar{C} - \bar{C}_{\Id},
    \label{eq:ergod_spec}
\end{equation}
where $\bar{C}_{\Id}$ is the equal power allocation solution. This quantity is independent of the additive channel contribution in~\eqref{eq:Ergodic_cap}. This implies that the physics of maximizing ergodic spectral efficiency is independent of the channel in uniform multipath environment, as long as the number of ports in the system is low~\cite{Gustafsson+Nordebo2007a}. Note that~\eqref{eq:ergod_spec} only contains a temporal average over the input signals, significantly simplifying its evaluation.

\section{Electric Field Integral Equation}
\label{app1}

All antenna structures in this paper are modeled as perfectly conducting bodies with the electric field integral equation (EFIE),~\cite{Chew+etal2008}, which relates tangential component of incident and scattered electric fields as
\begin{equation}
\label{eq:EFIE}
\hat{\boldsymbol{n}}\times\boldsymbol{E}_\mathrm{i} \left( \boldsymbol{r} \right) = - \hat{\boldsymbol{n}} \times \boldsymbol{E}_\mathrm{s} \left( \boldsymbol{r} \right),
\end{equation}
where $\boldsymbol{E}_\mathrm{i}$ is the incident field, $\boldsymbol{E}_\mathrm{s}$ is the scattered field
\begin{equation}
\boldsymbol{E}_\mathrm{s} \left( \boldsymbol{r} \right) = - \mathrm{j} k Z_0 \int\limits_{\varOmega} \mathbf{G} \left(\boldsymbol{r}, \boldsymbol{r} ' \right) \cdot \boldsymbol{J} \left(\boldsymbol{r} ' \right) \, \mathrm{d}S ',
\end{equation}
$Z_0$ is the impedance of vacuum, $k$ is wavenumber, and $\mathbf{G}$ denotes the free-space dyadic Green's function defined as
\begin{equation}
\mathbf{G} \left(\boldsymbol{r},\boldsymbol{r} '\right) = \left(\mathbf{1} + \frac{\nabla \nabla}{k^2} \right) \frac{\mathrm{e}^{-\mathrm{j} k | \mathbf{r} - \mathbf{r} ' | }}{4 \pi \left| \mathbf{r} - \mathbf{r} ' \right| }
\end{equation}
with~$\mathbf{1}$ being the identity dyadic.

By applying a suitable set of basis functions
\begin{equation}
\boldsymbol{J} ( \boldsymbol{r}) \approx \sum_{n=1}^{N} I_n \boldsymbol{\psi}_n (\boldsymbol{r})
\end{equation}
and the same set as testing functions (Galerkin method), the relation~\eqref{eq:EFIE} transforms into algebraic form
\begin{equation}
\mathbf{Z} \mathbf{I} = \mathbf{V},
\end{equation}
where the impedance matrix~$\mathbf{Z} = [Z_{nm}] \in \mathbb{C}^{N\times N}$ and excitation vector~$\mathbf{V} = [V_n] \in \mathbb{C}^{N \times 1}$ are defined element-wise as
\begin{equation}
Z_{nm} = - \mathrm{j} k Z_0 \int\limits_{\varOmega} \int\limits_{\varOmega} \boldsymbol{\psi}_n \left(\boldsymbol{r} \right) \cdot \mathbf{G} \left(\boldsymbol{r}, \boldsymbol{r} ' \right) \cdot \boldsymbol{\psi}_m \left(\boldsymbol{r} ' \right) \, \mathrm{d}S \, \mathrm{d}S ',
\end{equation}
and
\begin{equation}
V_n = \int\limits_\varOmega \boldsymbol{\psi}_n \left(\boldsymbol{r} \right) \cdot \boldsymbol{E}_\mathrm{i} \left( \boldsymbol{r} \right) \, \mathrm{d}S,
\end{equation}
respectively.

The real part~$\mat{R}$ of the impedance matrix~$\mat{Z} = \mat{R} + \mathrm{j} \mat{X}$ can be factorized as
\begin{equation}
\mat{R} = \mat{S}^\mathrm{T} \mat{S},
\label{eq:app1S1}
\end{equation}
where the matrix~$\mat{S}$ is defined element-wise as
\begin{equation}
S_{\alpha n} = k \sqrt{Z_0} \int\limits_{\varOmega} \boldsymbol{\psi}_n \cdot \mat{u}^{(1)}_\alpha (k \boldsymbol{r}) \, \mathrm{d}S
\label{eq:app1S2}
\end{equation}
with~$\mat{u}^{(1)}_\alpha$ being the regular spherical waves, see~\cite{Tayli2018} for details.

\section{Definition of Q-factor}
\label{app:Q}

The radiation Q-factor is defined as
\begin{equation}
Q_\mathrm{rad} = \dfrac{2\omega\max\left\{ W_\mathrm{m}, W_\mathrm{e} \right\}}{P_\mathrm{r}},
\label{eq:app1Q1}
\end{equation}
where the stored magnetic and electric energies are
\begin{subequations}
\begin{align}
\label{eq:app1Q2A}
\omega W_\mathrm{m} &= \dfrac{1}{8} \Jm^\herm \left( \omega\dfrac{\partial\mat{X}}{\partial\omega} + \mat{X} \right) \Jm, \\
\label{eq:app1Q2B}
\omega W_\mathrm{e} &= \dfrac{1}{8} \Jm^\herm \left( \omega\dfrac{\partial\mat{X}}{\partial\omega} - \mat{X} \right) \Jm,
\end{align}
\end{subequations}
the radiated power is defined as
\begin{equation}
P_\mathrm{r} = \dfrac{1}{2} \Jm^\herm \mat{R} \Jm,
\label{eq:app1Q3}
\end{equation}
and $\mat{Z} = \mat{R} + \mathrm{j} \mat{X}$ is the impedance matrix, see Appendix~\ref{app1}. For purposes of this paper, the stored energy matrix is introduced as
\begin{equation}
\mat{W}_\mathrm{x} = \dfrac{1}{2} \omega\dfrac{\partial\mat{X}}{\partial\omega},
\label{eq:app1Q4}
\end{equation}
and the radiation Q-factor is rewritten as
\begin{equation}
Q_\mathrm{rad} = Q + Q_\mathrm{t} = \dfrac{\omega\left( W_\mathrm{m} + W_\mathrm{e} \right)}{P_\mathrm{r}} + \dfrac{\omega|W_\mathrm{m} - W_\mathrm{e}|}{P_\mathrm{r}},
\label{eq:app1Q5}
\end{equation}
where the first part ($Q$) belongs to the antenna itself and the second part ($Q_\mathrm{t}$) belongs to a lumped element tuning an antenna to the resonance~\cite{Yaghjian+Best2005}. Generally, it holds that $Q_\mathrm{rad} \geq Q$, and $Q_\mathrm{rad} = Q$ only at the self-resonance of an antenna. In this paper, we use only the first part of~\eqref{eq:app1Q5}, $Q$, since it is a low-bound estimate of an achievable Q-factor irrespective of how a multi-port antenna is tuned.

\section{SVD of channel matrix}
\label{app:singval}
The optimal solution of~\eqref{eq:portCapOptdual} is found by writing the system on the form~\eqref{eq:waterfilling}. This is achieved through a change of variables. Define the matrix
\begin{equation}
    \mat{r}_\nu = \frac{1}{(\nu+1)}\left(\nu\mat{r} +Q^{-1}\mat{w}_\mathrm{x}\right) .
    \label{eq:rnu}
\end{equation}
This matrix is positive semi-definite for sufficiently large antenna sizes, and appropriate values of $\nu$. Positive semi-definite matrices can be decomposed with the Cholesky decomposition, $\mat{r}_\nu = \mat{b}^{\herm}\mat{b}$. With this decomposition the condition in~\eqref{eq:waterfilling} can be rewritten as,
\begin{equation}
    \medel{\portv^\herm\mat{r}_\nu\portv} = \Tr(\mat{r}_\nu\mat{a}) = \Tr(\mat{b}\mat{a}\mat{b}^{\herm}) = \Tr(\widetilde{\mat{a}}) ,
\end{equation}
where the cyclic permutation of the trace has been utilized and $\widetilde{\mat{a}}=\mat{b}\mat{a}\mat{b}^{\herm}$ is a change of variables. With this change of variables~\eqref{eq:portCapOptdual} is written as
\begin{equation}
\begin{aligned}
& \min\limits_\nu\,\max\limits_{\widetilde{\mat{a}}} &&  \log_2\det(\Id+\gamma\mat{s}\mat{b}^{-1}\widetilde{\mat{a}}\mat{b}^{-\herm}\mat{s}^{\herm}) \\
& \subto && \Tr(\widetilde{\mat{a}}) = 1. \\
\end{aligned}
\end{equation}
This is a problem that can be solved by water-filling~\cite{Paulraj+etal2003,Molisch2011}, as in~\eqref{eq:waterfilling}, with the channel matrix, $\mat{h}=\mat{s}\mat{b}^{-1}$. All that is required is the singular value decomposition of $\mat{h}$. The singular values can be calculated from the eigenvalues of the matrix times itself,
\begin{equation}
    \svd(\mat{h}) = (\eig(\mat{h}\mat{h}^\herm))^{1/2} = (\eig(\mat{s}\mat{r}_\nu^{-1}\mat{s}^\herm))^{1/2} .
    \label{eq:app_eigprob}
\end{equation}
The eigenvalues can be determined by utilizing the decomposition of the radiation matrix, $\mat{r}=\mat{s}^\herm\mat{s}$. Putting this into~\eqref{eq:app_eigprob},
\begin{multline}
    \eig(\mat{s}\mat{r}_\nu^{-1}\mat{s}^\herm) = \eig((\nu+1)\left(\nu +Q^{-1}\mat{s}^{-\herm}\mat{w}_\mathrm{x}\mat{s}^{-1}\right)^{-1}) \\
    = (\nu+1)\left(\nu +Q^{-1}\eig(\mat{s}^{-\herm}\mat{w}_\mathrm{x}\mat{s}^{-1})\right)^{-1}.
\end{multline}
The eigenvalues $\eig(\mat{s}^{-\herm}\mat{w}_\mathrm{x}\mat{s}^{-1})$ are determined from the generalized eigenvalue problem~\eqref{eq:eig_Emodes_port}. This gives the singular values~\eqref{eq:SVD_channel},
\begin{equation}
    \sigman^2=\dfrac{(1 + \nu)}{\emod/Q+\nu}.
\end{equation}

\bibliographystyle{IEEEtran}
\bibliography{total, additionsMIMOQ}

\begin{IEEEbiography}[{\includegraphics[width=1in,height=1.25in,clip,keepaspectratio]{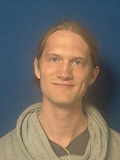}}]{Casimir Ehrenborg} 
(S'15) received his M.Sc. degree in engineering physics in 2014, and his Ph.D. degree in electrical engineering in 2019, from Lund University, Sweden. He is currently a postdoctoral fellow in the META research group at KU Leuven. In 2015, he  participated in and won the IEEE Antennas and  Propagation Society Student Design Contest for his body area network antenna design. In 2019, he was awarded the IEEE AP-S Uslenghi Letters Prize for best paper published in IEEE Antennas and Propagation Letters during 2018. His research interests include Spacetime metamaterials, MIMO antennas, physical bounds, and stored energy.
\end{IEEEbiography}

\begin{IEEEbiography}[{\includegraphics[width=1in,height=1.25in,clip,keepaspectratio]{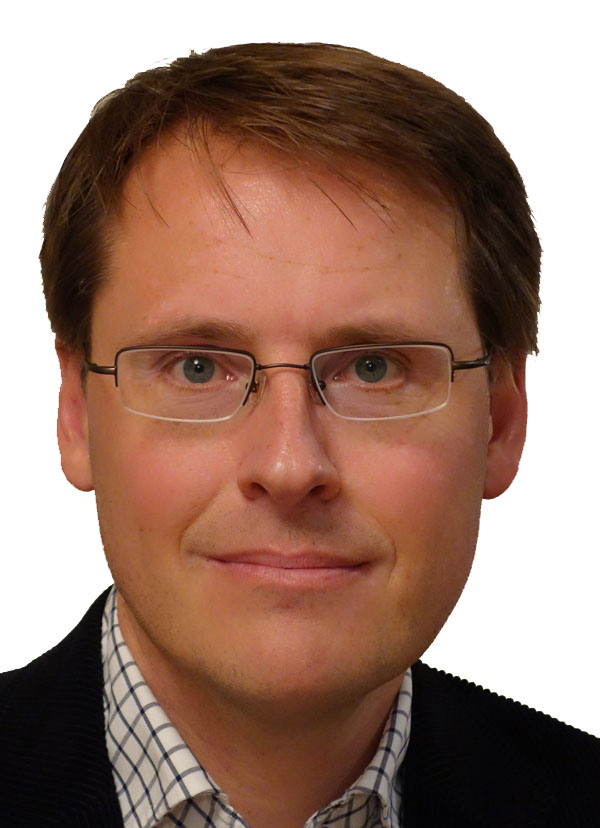}}]{Mats Gustafsson}
received the M.Sc. degree in Engineering Physics 1994, the Ph.D. degree in Electromagnetic
Theory 2000, was appointed Docent 2005, and Professor of Electromagnetic Theory 2011, all
from Lund University, Sweden.

He co-founded the company Phase holographic imaging AB in 2004. His research interests are in scattering and antenna theory and inverse scattering and imaging. He has written over 100 peer reviewed journal papers and over 100 conference papers. Prof. Gustafsson received the IEEE Schelkunoff Transactions Prize Paper Award 2010, the IEEE Uslenghi Letters Prize Paper Award 2019, and best paper awards at EuCAP 2007 and 2013. He served as an IEEE AP-S Distinguished Lecturer for 2013-15.
\end{IEEEbiography}

\begin{IEEEbiography}[{\includegraphics[width=1in,height=1.25in,clip,keepaspectratio]{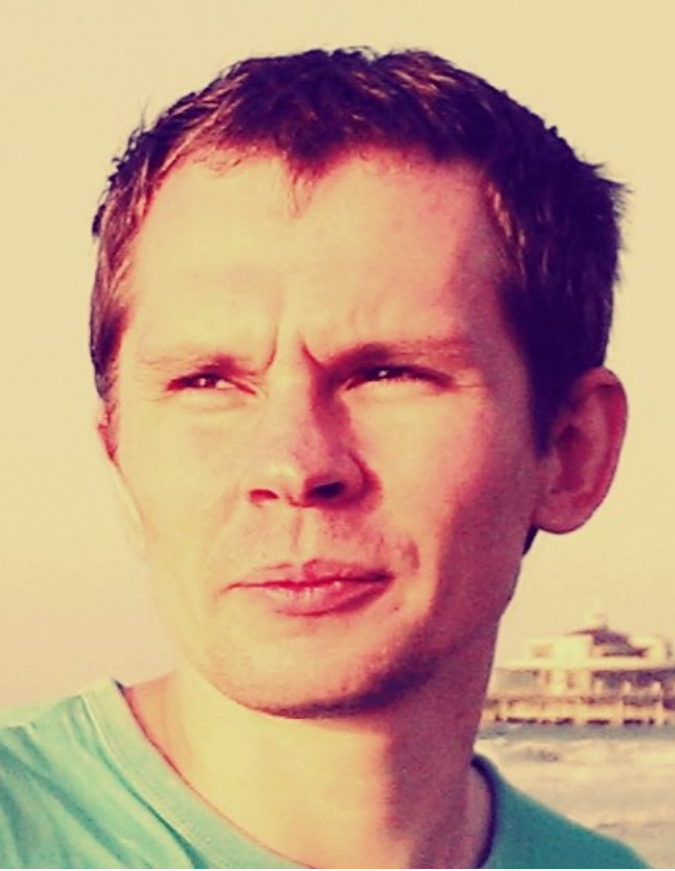}}]{Miloslav Capek}
(M'14, SM'17) received the M.Sc. degree in Electrical Engineering 2009, the Ph.D. degree in 2014, and was appointed Associate Professor in 2017, all from the Czech Technical University in Prague, Czech Republic.
	
He leads the development of the AToM (Antenna Toolbox for Matlab) package. His research interests are in the area of electromagnetic theory, electrically small antennas, numerical techniques, fractal geometry, and optimization. He authored or co-authored over 100~journal and conference papers.

Dr. Capek is member of Radioengineering Society, regional delegate of EurAAP, and Associate Editor of IET Microwaves, Antennas \& Propagation.
\end{IEEEbiography}

\end{document}